\documentclass[fleqn,12pt,twoside]{article}
\usepackage{espcrc1}

\usepackage{graphicx}

\newcommand{\AmS}{{\protect\the\textfont2
  A\kern-.1667em\lower.5ex\hbox{M}\kern-.125emS}}

\title{Finite Formation Time in Electro-Disintegration of Few-Body Nuclei}

\author{M. Braun \address[MCSD]{Department of High Energy Physics, St. Petersburg University, 198904 St. Petersburg, Russia},
C. Ciofi degli Atti\address
[MCSD]
{Department of Physics, University of Perugia an INFN Sezione di Perugia, 06100 Perugia, Italy},
        L. P. Kaptari\addressmark\thanks{On leave from Bogoliubov Laboratory of Theoretical Physics, 141980,
         JINR, Dubna, Russia.}
        and
        H. Morita\address[MCSD]{Faculty of Social Information, Sapporo Gakuin University, 11 Bunko-dai, Ebetsu, Hokkaido
        069-8555, Japan.}}
       
\begin{document}

\maketitle

\begin{abstract}
Inclusive and exclusive electro-disintegration of $^2H$, $^3He$ and $^4He$ are calculated within an approach where,
besides the final state interaction (FSI), also the finite formation
 time (FFT) of the hit hadron to its asymptotic physical state
is taken into account.
\end{abstract}

\section{Formalism} When a proton is knocked out off a nucleus in  quasi-elastic inclusive ($A(e,e')X$)  and 
exclusive 
($A(e,e'p)B$) processes at high values of the momentum transfer $Q^2$ , the Final State Interaction (FSI) can be described 
within the eikonal Glauber approach.
Recently   such an  approach to FSI
 has been extended
 so as to  take into account the virtuality of the hit  nucleon
  after $\gamma*$ absorption \cite{BCT}.
  It has been found that because of 
  the FFT   the hit  proton needs to reach 
  its asymptotic state, the FSI
   with the spectator $A-1$ nucleons becomes very  weak and vanishes
    in the asymptotic
   limit. Although the physical process underlying the vanishing
    of the FSI at large $Q^2$ is the
   same which appears
    in colour transparency phenomena, quantitative effects may largely 
   depend
    upon
   the details of the underlying theoretical model to treat FSI.
   In this contribution some  results of a systematic calculation of FSI and FFT effects
    in the  inclusive and exclusive processes off few-nucleon systems 
    will be presented. 
   The central quantity in our calculations is the  effective  nucleon  distorted momentum distributions
\begin{eqnarray}
n_{eff}(\vec{p}_m)  &=& \left |(2\pi)^{-3/2}\int d\vec{\xi} 
e^{-i\vec{p}_m\cdot\vec{\xi}} I(\vec{\xi})\right |^2
\end{eqnarray}
\noindent where ${\vec p}_m ={\vec q} - {\vec p}$\,  and\, ${\vec p}$\, 
are  the missing momentum and  the momentum of the detected proton, respectively, and
 $I(\vec{\xi})$  the distorted overlap between the 
ground state wave functions of nuclei $A$ and $(A-1)$, {\it viz}
\begin{eqnarray}
 I(\vec{\xi})     &=& \sqrt{A}\int 
   \psi_{A-1}^*(\xi_{A-1})S_G
   \psi_{A}(\xi_{A})\prod_{\xi_i}d\xi_i,
\end{eqnarray}
where $\xi_A$ and $\xi_{A-1}$ are the sets of Jacobi coordinates for nuclei $A$ and $A-1$, respectively.
\noindent If FFT effects are considered at 
$x\simeq 1$ ($x$ is  the Bjorken scaling variable), the Glauber operator $G(Ai) = 1-\theta(z_i-z_A)\Gamma(\vec{b}_A-\vec{b}_i)$
appearing in  $S_G = \prod_{i=1}^{A-1} G(Ai)$ can be replaced by \cite{BCT,HIKO}
\begin{eqnarray}
G(Ai)       = 1-{\mathcal{J}}(z_i-z_A)\Gamma(\vec{b}_A-\vec{b}_i), \qquad 
{\mathcal{J}}(z) = \theta(z)(1-exp(\frac{zxmM^2}{Q^2})), 
\label{eq:FFT}
\end{eqnarray}
where  $m$ the  nucleon mass, and  $M$
  the  average virtuality defined by $M^2 = (\overline {m^*}^2 - m^2$.
 Eq. \ref{eq:FFT} shows that at high values of  $Q^2$
 FFT effects reduce the  Glauber-type FSI, depending on the value of 
 $M$. At $x>1$ (the so called {\it cumulative} region), Eq. \ref{eq:FFT} does not
 hold and a simple prescription to deal with FFT effects cannot be given.  
 \section{Results} 
\begin{figure}[htbp]
      \centerline{
        \includegraphics[width=70mm, height=70mm]{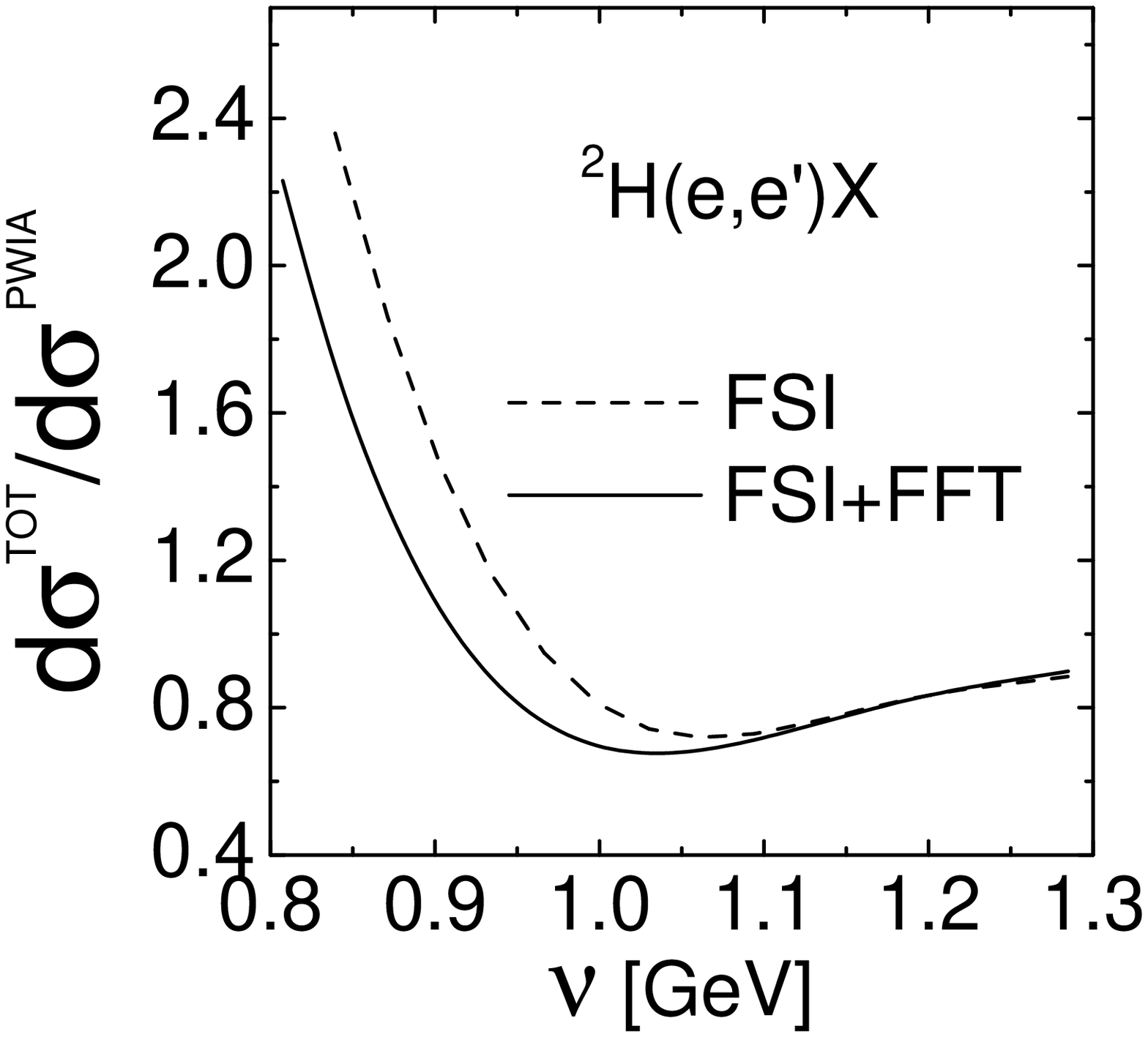}
        \hspace{15mm}
         \includegraphics[width=70mm, height=70mm]{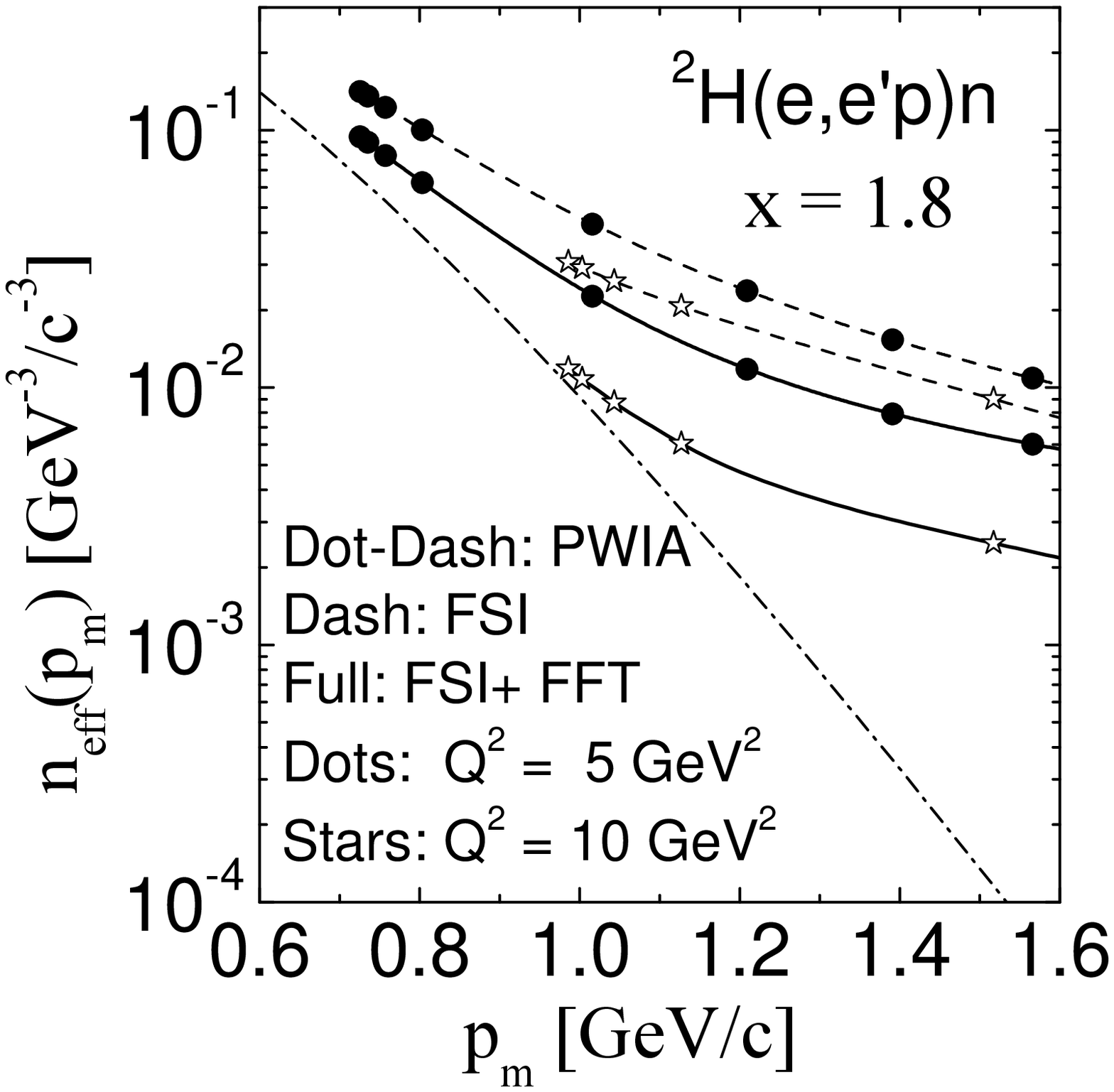}}   
      \vspace{-10mm}           
      \caption{{\it Left panel}: the ratio between the inclusive cross section
      $^2H(e,e')X$ including FSI (dashed line) and FSI+FFT (full line) to the PWIA cross section.
      The calculation corresponds to an initial electron energy $E_e=9.761 GeV$\,  and scattering angle $\theta_e=10^{o}$.
      The value of the energy transfer $\nu=1.3 \,GeV$ corresponds to $x=1$ and $\nu=0.83\,GeV$ to $x=1.71$
      (after Ref. \cite{BCK}). {\it Right panel}:  the $Q^2$-dependence of $n_{eff}$ at $x=1.8$ in the exclusive reaction
       $^2H(e,e'p)n$  (after Ref. \cite{BCK}).} 
 \vspace{-10mm}
 \label{Fig1}
\end{figure}

\begin{figure}[htbp]
\vspace{-5mm}
\centerline{
     \includegraphics[width=70mm,height=70mm]{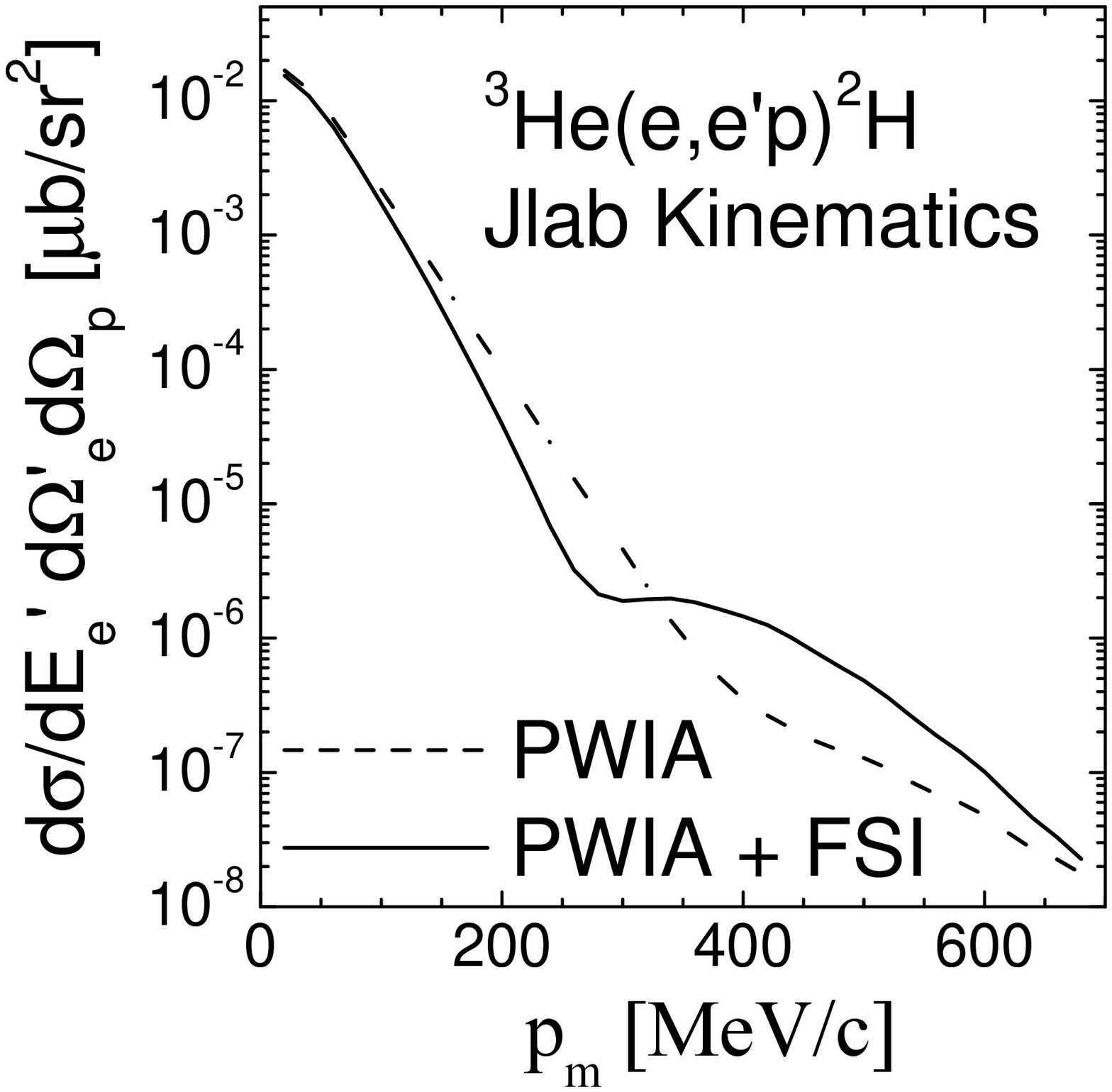}
     \hspace{10mm}
\includegraphics[width=70mm,height=70mm]{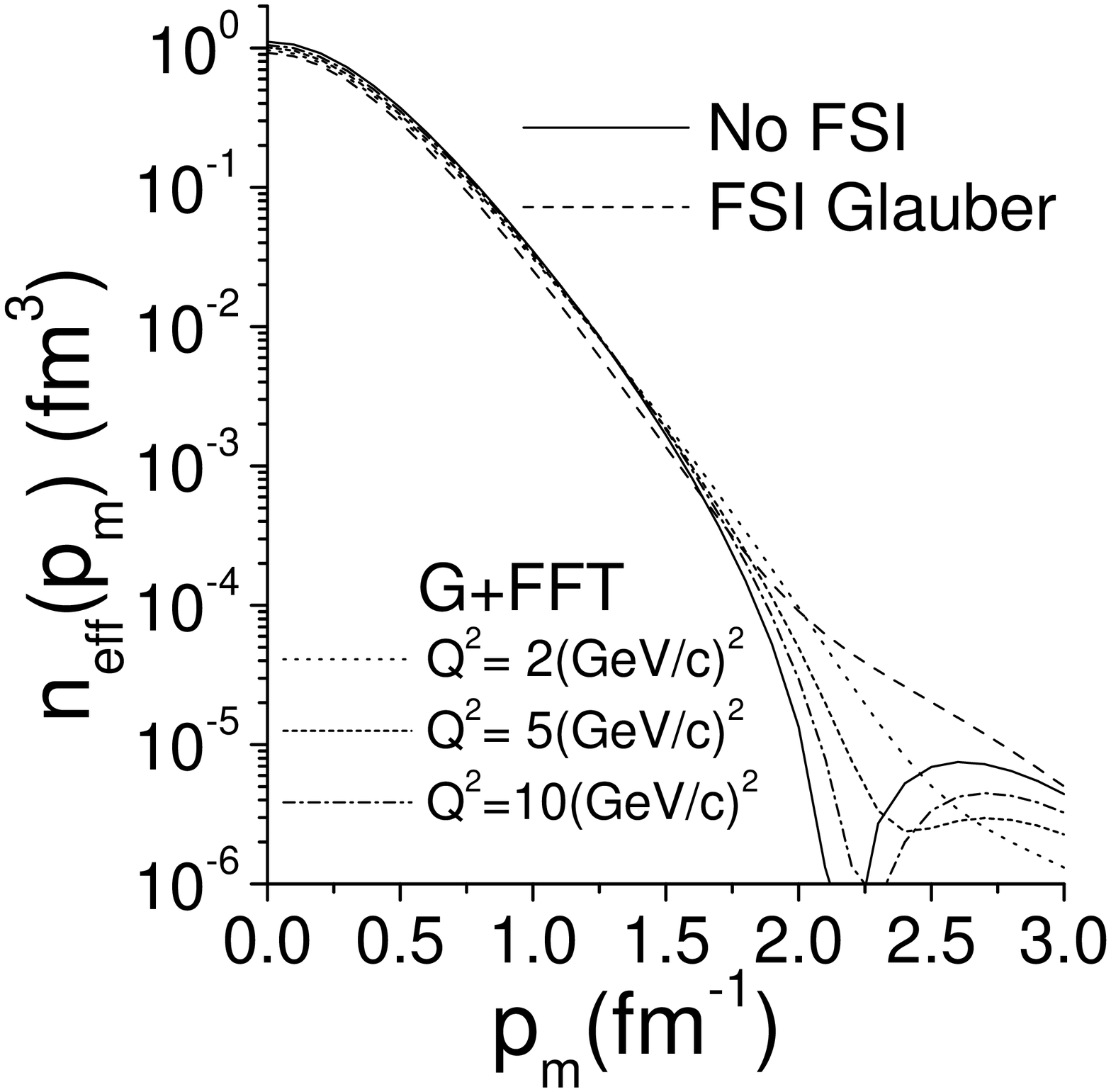}}
\vspace{-10mm}
      \caption{{\it Left panel}: the cross section of the process  $^3He(e,e'p)^2H$ in the kinematics of
      the Jlab experiment E-01020 \cite{E01020}. {\it Right panel}: the $Q^2$ dependence of $n_{eff}(\vec{p}_m)$
       for the process
      $^4He(e,e'p)^3H$ at $x=1$ and parallel kinematics. The full curve represents the PWIA result, 
      the dashed curve includes
      Glauber  FSI and the other curves include also FFT effects (after Refs.\cite{HIKO} and \cite{CKM})}
      \label{Fig2}
      \vspace{-5mm}
\end{figure}

Some results of our calculations, performed with realistic wave functions
 corresponding to the RSC and AV18 interactions, are shown  in 
 Fig. \ref{Fig1}
 and \ref{Fig2}.   It should be stressed that
         when calculations are performed in the relatively low $Q^2$
 kinematics of the recent TJLab experiments \cite{ulmer,89044,E01020},
 a good agreement with preliminary experimental data is observed. 
\vspace{-5mm} 
 
 \section{Summary and conclusions}The results we have exhibited show that: i) in inclusive and exclusive processes
 off $^2H$, FFT effects (color transparency effects) produce a non negligible contribution
 in the cumulative region ($x>1$) already at moderately values  of $Q^2$, whereas at $x=1$ they appear to contribute only
 at much higher values of $Q^2$; ii) the treatment of FSI effects within the Glauber approach can
 properly describe exclusive processes off $^2H$, $^3He$ and $^4He$ at the kinematics of present Jlab experiments;
 iii) the inclusion of FFT in the process  $^4$He(e,e'p)$^3$H \cite{HIKO}  predicts a 
clean and regular $Q^2$-dependence of FSI effects leading to their vanishing 
 at moderately large values of $Q^2$; such a prediction would be 
 validated by the experimental observation of a dip in the cross 
 section at $p_m \simeq 2.2 fm^{-1}$. In Ref. \cite{omar} a colour transparency model
 has been used  to  analyze the  same process obtaining 
  results which quantitatively differ from the ones we have obtained. Experimental data at 
  $Q^2 \,\simeq \, 10\, (GeV/c)^2$
  would resolve this important issue concerning the  propagation of hadrons in the medium.

\end{document}